\documentclass[aps,pra,twocolumn,showpacs,groupedaddress]{revtex4-1}

\usepackage[T1]{fontenc}
\usepackage{graphicx}
\usepackage{color,ulem}

\newcommand\forget[1]{}
\newcommand\comment[1]{}
\newcommand\ket[1]{|#1\rangle}
\newcommand\us{$\mu$}

\begin{document}


\title{Confined quantum Zeno dynamics of a watched atomic arrow}

\author{Adrien Signoles}
\affiliation{Laboratoire Kastler Brossel, ENS, UPMC-Paris 6, CNRS, 24 rue Lhomond, 75005 Paris, France}
\affiliation{Coll\`ege de France, 11, place Marcelin Berthelot, 75231 Paris Cedex 05, France}
\author{ Adrien Facon}
\affiliation{Laboratoire Kastler Brossel, ENS, UPMC-Paris 6, CNRS, 24 rue Lhomond, 75005 Paris, France}
\affiliation{Coll\`ege de France, 11, place Marcelin Berthelot, 75231 Paris Cedex 05, France}
\author{Dorian Grosso}
\affiliation{Laboratoire Kastler Brossel, ENS, UPMC-Paris 6, CNRS, 24 rue Lhomond, 75005 Paris, France}
\affiliation{Coll\`ege de France, 11, place Marcelin Berthelot, 75231 Paris Cedex 05, France}
\author{ Igor Dotsenko}
\affiliation{Laboratoire Kastler Brossel, ENS, UPMC-Paris 6, CNRS, 24 rue Lhomond, 75005 Paris, France}
\affiliation{Coll\`ege de France, 11, place Marcelin Berthelot, 75231 Paris Cedex 05, France}
\author{Serge Haroche$^*$}
\affiliation{Laboratoire Kastler Brossel, ENS, UPMC-Paris 6, CNRS, 24 rue Lhomond, 75005 Paris, France}
\affiliation{Coll\`ege de France, 11, place Marcelin Berthelot, 75231 Paris Cedex 05, France}
\author{ Jean-Michel Raimond}
\affiliation{Laboratoire Kastler Brossel, ENS, UPMC-Paris 6, CNRS, 24 rue Lhomond, 75005 Paris, France}
\affiliation{Coll\`ege de France, 11, place Marcelin Berthelot, 75231 Paris Cedex 05, France}
\author{Michel Brune}
\author{S\'ebastien Gleyzes}
\affiliation{Laboratoire Kastler Brossel, ENS, UPMC-Paris 6, CNRS, 24 rue Lhomond, 75005 Paris, France}
\affiliation{Coll\`ege de France, 11, place Marcelin Berthelot, 75231 Paris Cedex 05, France}



\date{\today}


\maketitle

{\bf
In a quantum world, a watched arrow never moves. This is the Quantum Zeno Effect \cite{Misra1977}. Repeatedly asking a quantum system ``are you still in your initial state ?'' blocks its coherent evolution through measurement back-action. Quantum Zeno Dynamics (QZD) \cite{Facchi2002,Facchi2008} leaves more freedom to the system. Instead of pinning it to a single state, it sets a border in its evolution space. Repeatedly asking the system ``are you beyond the border ?'' makes this limit impenetrable. Since the border can be designed by choosing the measured observable, QZD allows one to tailor dynamically at will the system's Hilbert space. Recent proposals, particularly in the Cavity Quantum Electrodynamics (CQED) context \cite{Raimond2010,Raimond2012}, highlight the interest of QZD for quantum state engineering tasks \cite{Wang2008,Maniscalco2008,Rossi2009,Shao2009,Chandrashekar2010,Shi2012}, which are the key to quantum-enabled technologies and quantum information processing. We report the observation of QZD in the 51-dimension Hilbert space of a large angular momentum $J=25$. Continuous selective interrogation limits the evolution of this angular momentum to an adjustable multi-dimensional subspace. This confined dynamics leads to the production of non-classical `Schr\"odinger cat' states \cite{ETQ,Vlastakis2013}, quantum superpositions of angular momentums pointing in different directions. These states are promising for sensitive metrology of electric and magnetic fields. This QZD approach could also be generalized to cavity and circuit QED experiments \cite{Raimond2010,Raimond2012,Vlastakis2013}, replacing the angular momentum by a photonic harmonic oscillator.
} 

Quantum Zeno dynamics modifies the classical motion of a system by its observation in a quantum context \cite{Raimond2010,Raimond2012,Rossi2009,Wang2008,Maniscalco2008,Shao2009,Chandrashekar2010,Shi2012}. However, an actual projective quantum measurement is not mandatory, and QZD can be equivalently attained by performing a pulsed unitary acting only on the states at the border ("Bang Bang" control) or even by applying a strong continuous coupling to these states. This has been predicted theoretically \cite{Facchi2008} and verified in a recent experiment \cite{Schafer2014}. In that experiment, however, the evolution of the system is restricted to a dimension 2 subspace. The dynamics is simply that of a spin 1/2, and do not exhibit the most striking features of QZD \cite{Raimond2010}. 

In this Letter we implement QZD in a large atomic angular momentum $J=25$ (`spin' or top), represented as an arrow pointing on a generalized Bloch sphere. In the 51-dimensional Hilbert space, we isolate tailorable multi-dimensional manifolds. We show how QZD induces a very non-classical dynamics inside the Zeno subspace, leading to the generation of Schr\"odinger cat spin states \cite{ETQ}, in which the arrow points at the same time in two different directions. As spin-squeezed states \cite{Ockeloen2013}, which are the focus of an intense attention, these cat states lead to quantum-enabled metrological applications \cite{Vlastakis2013}.

The angular momentum projection on the polar axis of the generalized Bloch sphere is quantized, taking the values  $J-k$, with $k=0\ldots 2J$ (the corresponding eigenstates being $|J,J-k\rangle$). The dynamical evolution from the initial state $|J,J\rangle$ (North pole of the Bloch sphere) is induced by a resonant field driving transitions between these eigenstates. In classical terms, it corresponds to a rotation of the arrow along a meridian from the North to the South pole and back. In quantum terms, at each stage of the rotation, the system is in a spin coherent state \cite{Arecchi1972}, superposition of  $|J,J-k\rangle$s, the average value of $J-k$ coinciding with the projection of the arrow on the polar axis. 

Repeatedly measuring the value of this projection would freeze the rotation, merely realizing the quantum Zeno effect. Here, instead, we implement the QZD by applying continuously a selective unitary evolution addressing only one of the $|J,J-k\rangle$ states. This state corresponds to a well-defined `limiting latitude' on the Bloch sphere. The spin is forbidden to cross the limiting latitude and its motion remains confined on the North polar cap \cite{Raimond2012}.
 
This confined motion is nontrivial. As the rotating spin reaches the limiting latitude crossing point, it suddenly vanishes and reappears at a point on the limiting latitude with opposite longitude (inversion of the spin's azimuthal phase). The rotation then resumes towards the North pole. The complete, smooth rotation of the classical dynamics is  interrupted by sudden phase inversions and replaced by a confined motion on the polar cap bounded by the limiting latitude. Caught at the phase inversion time,  the spin is transiently in a quantum superposition of two spin coherent states pointing along opposite longitudes, a cat state.

This confined evolution is similar to that predicted for QZD in the cavity or circuit quantum electrodynamics context \cite{Raimond2010,Raimond2012}. The dynamics of an angular momentum near the North pole of the Bloch sphere is analogous  to that of a one-dimensional field oscillator, $k$ playing the role of the photon number \cite{ETQ}. In this analogy, the polar cap of the Bloch sphere becomes the phase plane spanned by the field quadratures. Our experiment can thus be viewed as a quantum simulation of the CQED version of QZD. 

\begin{figure}
 \includegraphics[width=\linewidth]{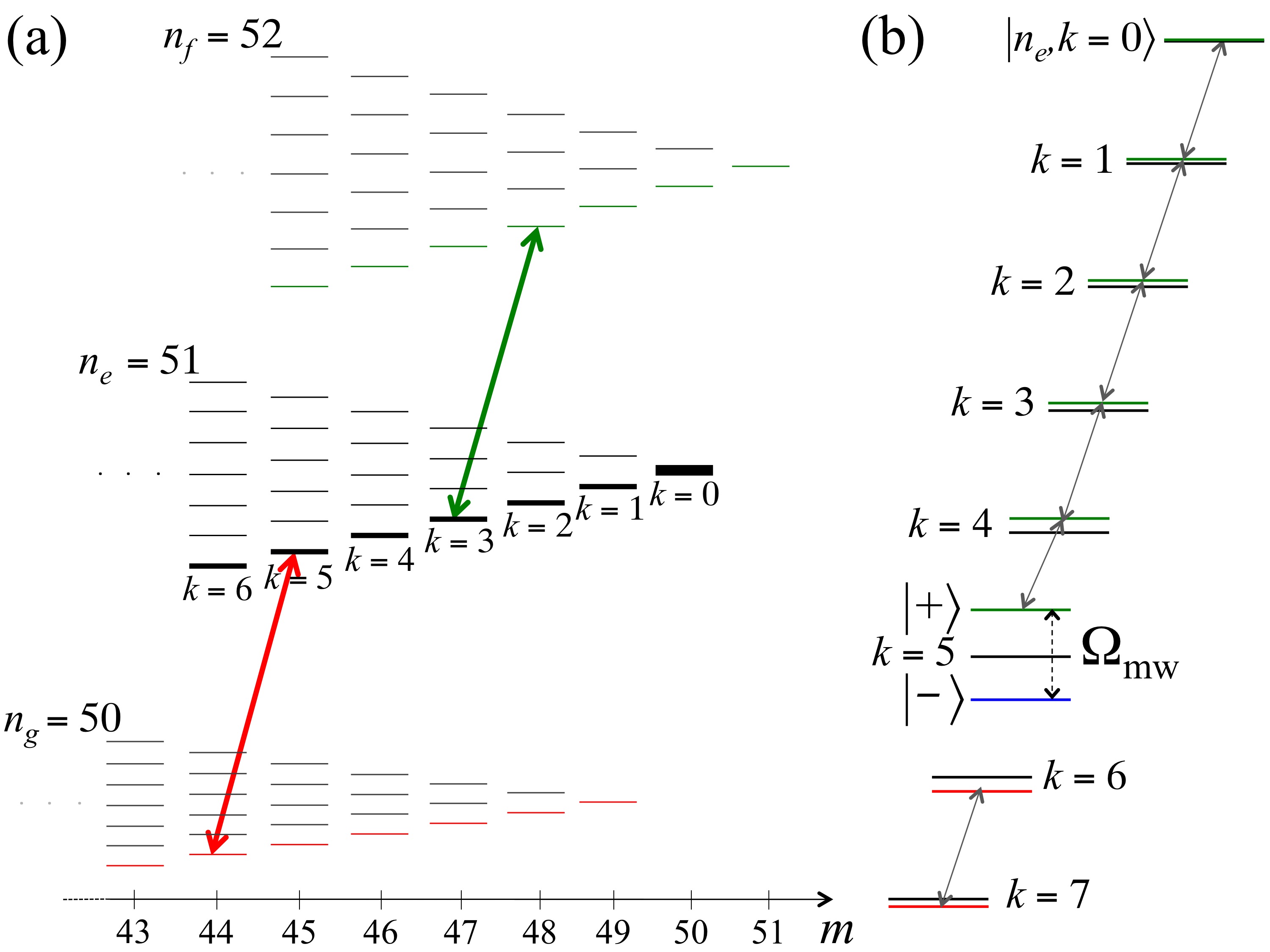}
  \caption{{\bf Rydberg energy levels } (a) Stark manifolds with principal quantum numbers $n_g=50$, $n_e=51$ and $n_f=52$ (not to scale). In each manifold, the levels, sorted by their magnetic quantum number $m$, form a triangular pattern. The thick lines represent the spin states ladder  coupled to the circular state $\ket{n_e,0}$ (thickest line) by the $\sigma_+$ RF field. The green and red arrows show the transitions resonant with the probe pulse for $k_p=3$ and with the Zeno MW for $k_z=5$ respectively. (b) Spin states without (black lines) and with (colored lines) the Zeno MW.  The Zeno field splits the $|n_e,k_z\rangle$ state in two dressed states, $\ket +$ (green line) and  $\ket -$ (blue line), separating the Hilbert space in two subspaces ${\cal H_N}$ and ${\cal H_S}$  (spanned by green and red states respectively). The RF field induces transitions within these subspaces (tilted arrows) but is unable to connect them through the gap $\Omega_{mw}$ opened by the Zeno MW. The Rabi splitting and the light shifts are exaggerated for clarity.
 }\label{fig-schematic}
\end{figure}

The spin $J=25$ is implemented in a subspace of the Stark manifold of a Rydberg atom. The interest of coherent manipulations of Rydberg manifolds has already been demonstrated in pioneering experiments on coherent wave packet dynamics  \cite{Bromage1999,Ahn2000,Dunning2009}. We take advantage of the versatility of this system to demonstrate here a new quantum feature. Figure 1(a) sketches parts of three adjacent Rydberg manifolds \cite{GallagherBook} (principal quantum numbers $n_f=52$, $n_e=51$ and $n_g=50$) in a static electric field $\mathbf{F}$, defining the quantization axis $Oz$. The eigenstates are sorted in columns according to their magnetic quantum number $m$ (selected to be positive). The circular state \cite{Hulet1983,Cheng1994} in the $n_e$ manifold (thickest line) has the maximum allowed $m=n_e-1$ value. A $\sigma_+$-polarized radio-frequency (RF) field couples it to a ladder of nearly equidistant levels (thick lines). The transitions between adjacent ladder states are at the Stark angular frequency  $\omega_a=(3/2)n_e ea_0 F/\hbar$ within small second order corrections in $F$  ($a_0$: Bohr radius, $e$: charge quantum). Since the atom is prepared initially in the circular state, the other levels in the manifold are not populated by the RF-induced dynamics and are ignored. The atom evolves within a ladder of 51 levels, $|n_e,k=0\rangle\ldots |n_e,k=50\rangle$, where $|n_e,k=0\rangle$ is the circular state.

The coherent evolution induced by the RF field is ruled by the Hamiltonian \cite{EnglefieldBook}:
\begin{equation}
 \hat V  = \frac{\hbar\Omega_{rf}}{2}\sum_{k}\sqrt {(k+1)(n_e-k-1)} |n_e,k+1\rangle\langle n_e,k|+\mbox{h.c.}\  .\label{eqn-V}
\end{equation}
This Hamiltonian describes the rotation of a $J=25$ angular momentum at a Rabi frequency  $\Omega_{rf}$ \cite{Arecchi1972}, with the correspondence $|n_e,k\rangle\rightarrow|J,J-k\rangle$. 

The atomic state, driven by the RF, moves down and up the ladder, while the equivalent angular momentum rotates around a meridian of its Bloch sphere. We observe this rotation by applying the resonant RF for a time $t_1$  and we measure the populations of $|n_e,k\rangle$ as a function of $t_1$ by field-ionization (See supplementary information). However, field-ionization does not resolve directly the $\ket {n_e, k}$ states. Hence the population $P(k_p,t_1)$ of $\ket{n_e,k_p}$ is measured by applying, before field-ionization, a resonant $\pi$-microwave (MW) pulse tuned to the  $\ket{n_e,k_p}\rightarrow\ket{n_f,k_p}$ transition [green arrow on Fig. 1(a)]. This pulse does not address the levels $\ket {n_e, k\not = k_p}$, due to the difference between the linear Stark frequencies in adjacent manifolds. Field-ionization selectively measures the population of $\ket {n_f, k_p}$, equal to $P(k_p,t_1)$ within the $\pi$-pulse transfer efficiency, $\eta_{k_p}\sim0.9$ (Supplementary information).

Figure 2(a) shows $P(k,t_1)$ ($k=0\ldots 5$) versus $t_1$, for $F=2.35$~V/cm corresponding to $\omega_a/2\pi=230.15$ MHz and $\Omega_{rf}/2\pi=152\pm 4$~kHz. The conspicuous cascade down the state ladder reveals the spin rotation. The insets show snapshots of the population distribution in the ladder levels. Data are in excellent agreement with the theoretical expectations for a rotating spin coherent state \cite{Mogensen1995}.

\begin{figure}
    \includegraphics[width=\linewidth]{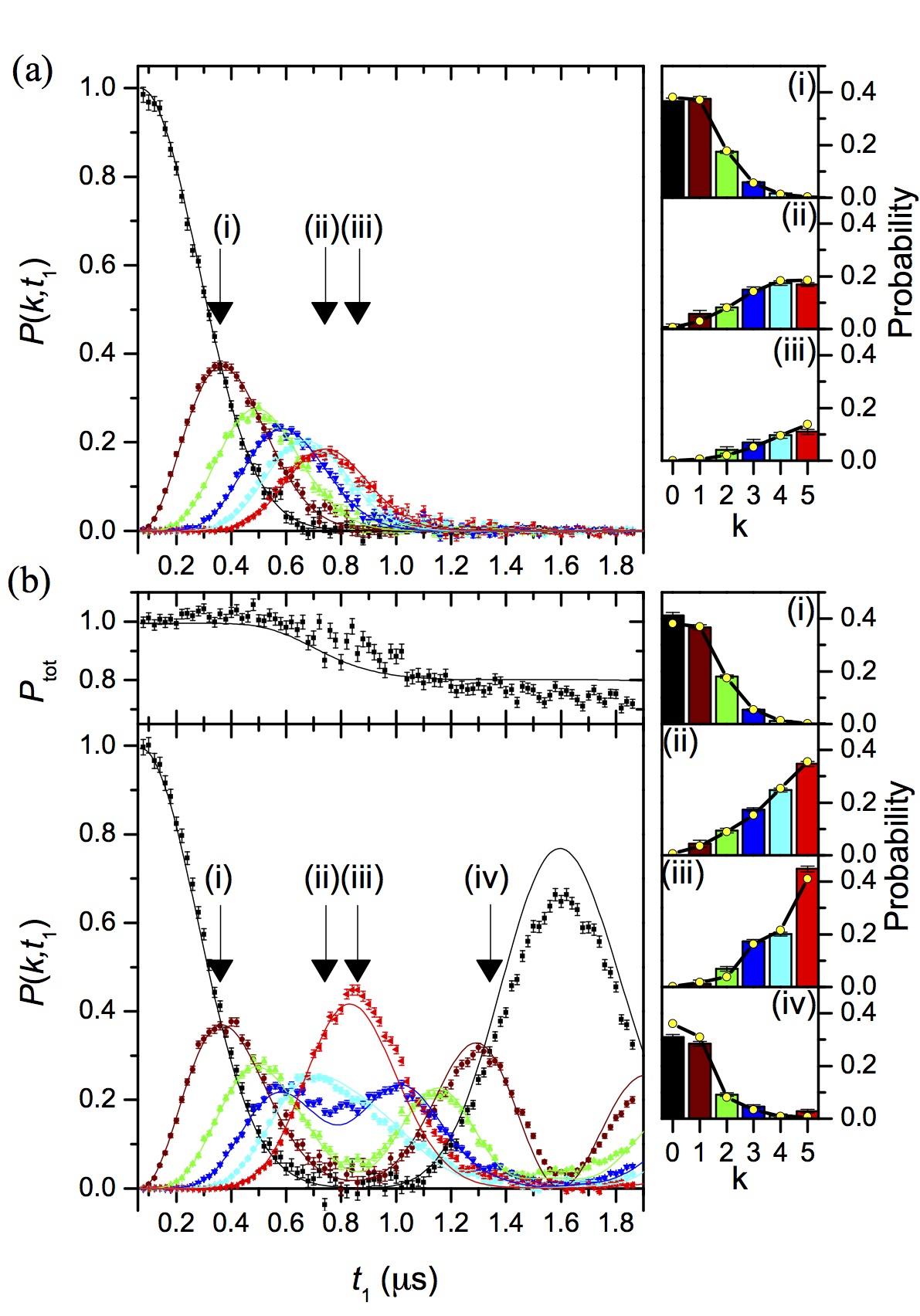}
  \caption{ {\bf   Evolution of the spin state populations.} (a) $P(k,t_1)$ for $k\le 5$ in a free RF-induced rotation (no Zeno MW applied). The points are experimental with statistical error bars. The solid lines correspond to the theoretical expectation for a spin coherent state rotating at the fitted frequency $\Omega_{rf}$. The insets define the color code for the different $k$ values and present the experimental histograms of $P(k,t_1)$ for three $t_1$ values (arrows in the main frame), together with the expected spin coherent state distribution (solid lines with yellow dots).  (b) $P(k,t_1)$ for $k\le 5$ in a QZD with  $k_z=5$. The top frame gives the total population $P_{tot}$ of ${\cal H_N}$. The lines result from the complete numerical simulation of the experiment (Methods). The insets give the observed $P(k,t_1)$ distribution at four $t_1$ values (arrows in the main frame) together with the numerical predictions (solid lines with yellow dots). }  \label{fig-schematic}
\end{figure}

To induce the QZD, we continuously interrogate the atom by selectively addressing one of the spin states with a `Zeno' c.w.  MW field resonant on the transition $\ket {n_e, k_z}\rightarrow\ket {n_g, k_z} $ [red arrow on Fig. 1(a)]. For levels $\ket {n_e, k\not =k_z}$, this MW is non-resonant and produces only small light shifts. For $k=k_z$, the Zeno MW admixes $\ket {n_g, k_z}$ with  $\ket {n_e, k_z}$ replaced by a pair of dressed states, $\ket{\pm}$, separated by $\Omega_{mw}$ (dynamical Stark splitting).  

The resulting level ladder is sketched in Fig.1(b). The $\sigma_+$ transitions within the subspace $\{ \ket {n_e, 0},  \ket {n_e, 1},\ldots  \ket {n_e, k_z-1},\ket{+}\}$ [arrows in Fig. 1(b)] are nearly degenerate at the frequency $\omega_a$. The Zeno MW dressing opens, between $\ket +$ and $\ket -$,  a gap wider than the coupling matrix element $\sim\sqrt{n_e}\Omega_{rf}$ of $\hat V$ between the spin states. It  makes it nearly impossible for the RF drive to induce, in an evolution from $\ket {n_e, 0}$,  transitions towards states below $\ket +$. The population of $\ket -$ is negligible and this state can be disregarded in the discussion.  Moreover, after an appropriate adiabatic switching-off of the Zeno MW (Supplementary information), $\ket +$ is mapped onto  $|n_e,k_z\rangle$ ($\ket -$, being mapped onto $\ket{n_g,k_z}$). The QZD thus splits the angular momentum Hilbert space into ${\cal H}_N$, made up of the $k_z+1$ levels with $k\le k_z$ close to the North pole of the Bloch sphere, and the complementary southern subspace ${\cal H}_S$ ($k>k_z$).

After a RF-induced QZD lasting a time $t_1$, we probe the level populations in  ${\cal H}_N$. We adiabatically switch off the Zeno MW and measure $P(k,t_1)$. Figure 2(b) presents the results of this procedure for $k_z=5$ and $\Omega_{mw}/2\pi=3.4$~MHz. The state distribution now bounces off a `wall' at $k=k_z+1$ and nearly returns into the initial state after 1.6$\ \mu$s. This dynamics is drastically different from the runaway process observed without Zeno MW [Fig. 2(a)]. It is in excellent agreement with a complete numerical simulation of the experiment based on the independently measured experimental parameters (Supplementary information).

The top frame in Figure 2(b) shows the total population detected in $\cal H_N$. It drops by $\sim 25$\% at the bouncing time. This loss is mainly due to a residual transfer into $\cal H_S$ through the Zeno barrier. The insets show the histograms of $P(k,t_1)$ at four different times.  They are radically different from those obtained without QZD [Fig. 1(a)]. We can clearly see that the level population at the bouncing time is no longer that of a coherent spin state. Not only QZD restricts the evolution to a subspace of 5 states instead of 51, but the dynamics itself exhibits striking non-classical features. 

We get a clearer picture of this dynamics by a direct measurement of the spin's $Q$-function \cite{Agarwal1998}, transposing to spin systems the quantum optics Husimi distribution. It is defined on the Bloch sphere as $Q(\theta,\phi)=(2J+1)/(4\pi)\;\langle n_e, 0|R^\dagger (\theta,\phi)\rho R(\theta,\phi) |n_e,0\rangle $, where $\rho$ is the angular momentum density operator and $R$ the rotation along a meridian of the Bloch sphere bringing the North pole in the direction defined by the polar angles $\theta$ and $\phi$. 

Determining $Q$ thus amounts to measuring the population in $|n_e,0\rangle$ after rotating the state by means of a resonant RF pulse whose duration $t_2$ controls $\theta$ and whose adjustable phase controls $\phi$. We perform this rotation with a RF power much larger than that used for the QZD (coupling $\sqrt n \Omega'_{rf}=2\pi \cdot 6.3\mathrm{ MHz} >\Omega_{mw}$). It couples ${\cal H}_N$ and $\cal H_S$ even in the presence of the Zeno MW (Supplementary information).

\begin{figure*}
\includegraphics[width=\linewidth]{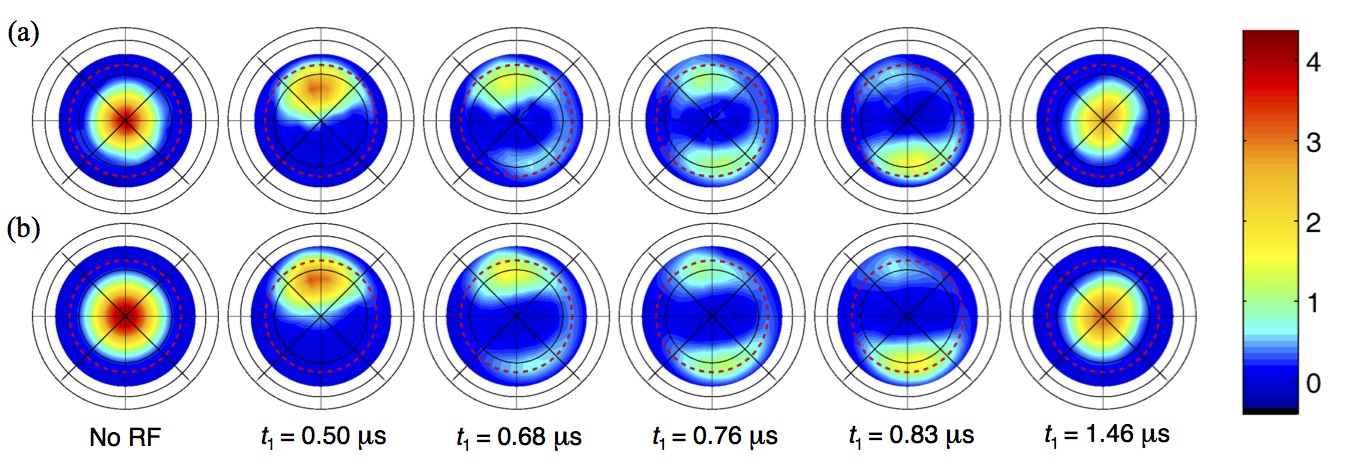}\\
  \caption{ {\bf Evolution of the $Q$-function under QZD.} (a) Measured $Q(\theta,\phi)$ functions for $k_z=4$ on the Bloch sphere, represented in polar projection. The black thin lines indicate 3 parallels separated by 30 degrees and 8 meridians. The data are linearly interpolated from 96 measurements points at different $(\theta,\phi)$ values spanning the North polar cap. Left frame: initial $|n_e,k=0\rangle$ state (no RF rotation). The other frames correspond to increasing times $t_1$ from left to right. The motion is confined by Zeno dynamics inside the limiting latitude (dashed red circle).  (b) Corresponding numerical predictions.}  \label{fig-Q}
  \end{figure*}

Figure 3(a) shows six snapshots of $Q$ for $k_z=4$ and $\Omega_{mw}/2\pi=3.08\pm 0.11$~MHz. The initial $|n_e,0\rangle$ state has a Gaussian $Q$-function centered at the North pole, which first moves, upwards in Fig. 3, towards the limiting latitude (dashed red line). It then splits in two components with opposite azimuthal phases. The upper one rapidly decreases, while the lower grows. At $t_1=0.76$~$\mu$s  the two peaks are balanced (fourth frame). After the phase inversion, the $Q$-function is mainly located in the lower part of the limiting latitude and resumes it motion towards the North pole, reached again at $t_1=1.46$ $\mu$s (last frame). Figure 3(b) presents the results of the full numerical simulation.  The excellent agreement between simulation and experiment confirms our understanding of the system Zeno dynamics and of the spin state measurement process.

At $t_1=0.76$~$\mu$s, we expect the system to be in a superposition of two spin coherent states with opposite azimuthal phases. However, the coherence of this superposition is not conspicuous in the $Q$-function. In order to get this information, we reconstruct the full angular momentum density matrix, $\rho$, at this time through a Maximum Likelihood method \cite{Lvovsky2004}. It is based on the measurement of the population of several levels after adjustable RF-induced rotations and adiabatic switching-off of the Zeno MW (Methods).

Figure 4(a) shows, on the Bloch sphere, the corresponding angular momentum Wigner function \cite{Dowling1994} $W(\theta,\phi)$ at $t_1=0.76$~$\mu$s. As in the quantum optics context, negative values for this quasi-probability distribution are an unambiguous indication of the state non classicality. We observe two positive maxima near the limiting latitude. They correspond to the two spin coherent state-like components pointing towards opposite azimuthal phases at the phase inversion time. In between, the interference fringes and their negativities give a vivid evidence that we prepare a genuine quantum superposition of two distinct mesoscopic spin states, a cat sate. These interference patterns cannot be observed when the Zeno subspace is only of dimension 2 \cite{Schafer2014}.  Figure 4(b) presents the simulated Wigner function taking into account the exact Hamiltonian of the system and all the known imperfections. Experiment and simulation are in excellent agreement (mutual fidelity 0.93).  The measured state has a purity Tr$\rho^2=0.75$ (simulation : 0.91). It is limited by static electric field inhomogeneities.

\begin{figure}
 \includegraphics[width=.8\linewidth]{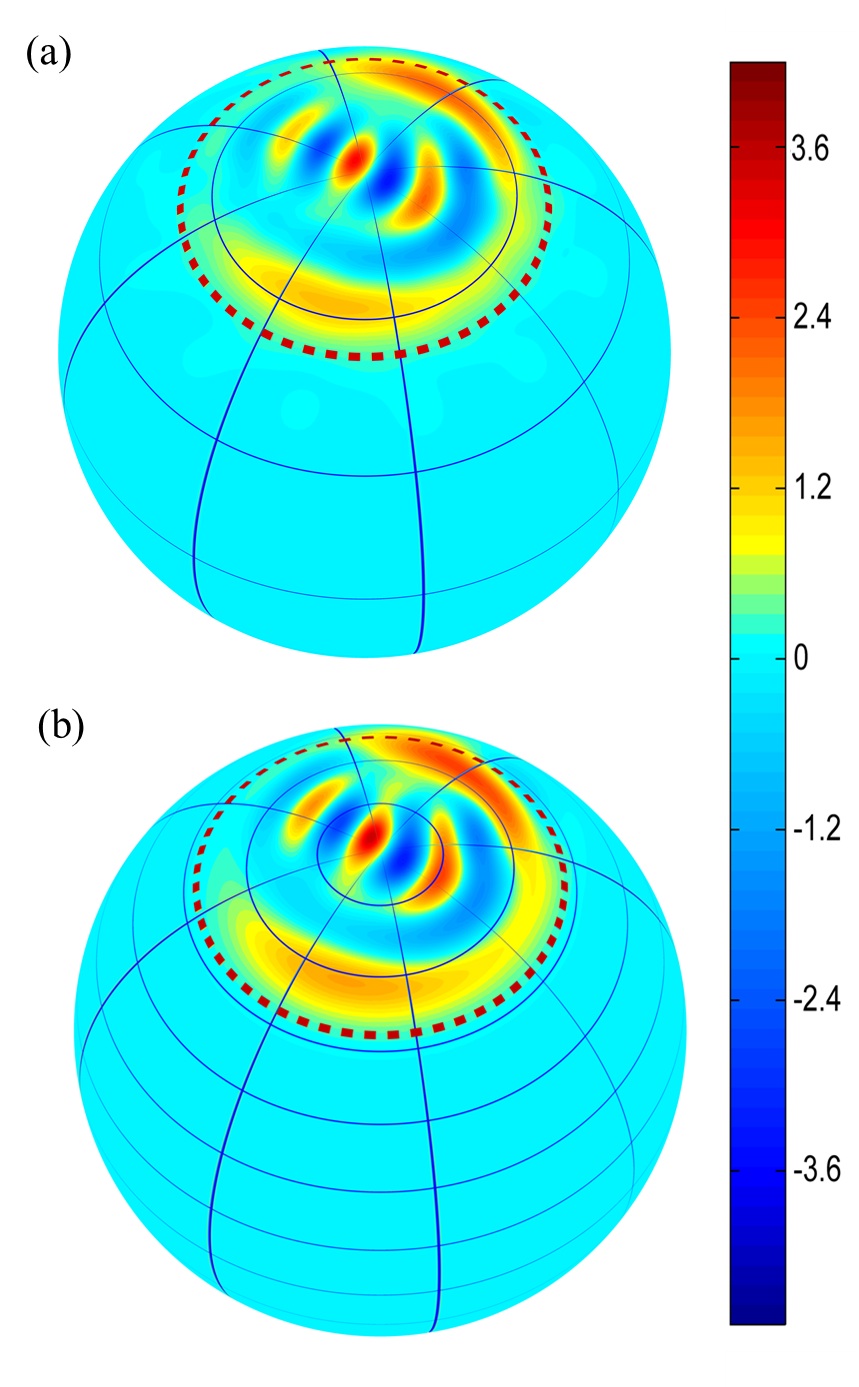}
  \caption{{\bf Wigner function of the spin cat state} (a) Experimental Wigner function, $W(\theta,\phi)$, obtained from the reconstructed density matrix $\rho$ of the spin after a QZD at the phase inversion time $t_1=0.76$ \us s (corresponding to the fourth frame in Fig.3).  The interference pattern between the two classical components reveals the non-classical coherence of this state. (b) Result of the numerical simulation of the experiment. The fidelity of the calculated density matrix $\rho_c$ with $\rho$ is Tr$^2(\sqrt{\sqrt\rho\rho_c\sqrt\rho})=$  0.93.}  \label{fig-Q}
\end{figure}

This experiment demonstrates the implementation of QZD in a Hilbert space large enough to allow us to generate mesoscopic superposition states. This is an important step towards quantum control through Hilbert space engineering.  It has been shown that the quantum control of the massively multi-level Rydberg states structure leads to important applications to state tailoring \cite{Bromage1999,Dunning2009} and quantum information \cite{Ahn2000}. The QZD opens an easily tailorable route towards the generation of such states. Moreover, the concepts and techniques used here are of general interest \cite{Rossi2009,Wang2008,Maniscalco2008,Shao2009,Chandrashekar2010,Shi2012} and could be applied, for instance, to superconducting qubits in circuit QED \cite{Vlastakis2013}, with direct applications to quantum information processing \cite{Massar2003}. 

We plan to extend these experiments to larger $k_z$ values and generate larger Schr\"odinger cat states, opening the way to metrology beyond standard quantum limit. The fast oscillations of the Wigner function near the north pole makes the measurement of its value a signal which is very sensitive to a small rotations \cite{Vlastakis2013}. Such states could therefore be used as very sensitive probe of small static magnetic or electric fields. 

We also plan to investigate engineered decoherence \cite{Poyatos1996}, through the application of a controlled electric field noise. The rich level structure of the Rydberg manifolds opens the way to the implementation of decoherence-free qubits through level-dressing schemes \cite{Aharon2013}. Furthermore, the atomic state could be mapped on that of a high-$Q$ cavity by tuning selected transitions in resonance by means of the Stark effect. The realization of a few qubits processor with a single multilevel atom in a cavity is within reach.


\section{Methods}

The atoms are produced by excitation of a thermal Rubidium beam. Two electrodes $A$ and $B$ facing each other (diameter 60 mm) produce the directing electric field $\mathbf{F}$ along $Oz$. The gap between $A$ and $B$ is surrounded by four independent electrodes, on which we apply RF signals to produce $\sigma_+$ fields with tunable phase and amplitude.

The atomic sample (1.5 atoms on the average) is prepared in the circular Rydberg state by pulsed laser excitation followed by an RF-induced adiabatic rapid passage through the spin ladder.  The whole process is completed in 5.6~\us s. Doppler-selective laser excitation addresses atoms with a velocity  $v=254\pm4$~m/s. 

The residual static field inhomogeneities and the atomic motion limit the useful observation time. The experimental sequence duration is 35 \us s, out of which at most 3 \us s are devoted to the QZD itself. The sequence includes the state preparation, the QZD, the RF rotation used for state reconstruction, the adiabatic Zeno MW switching-off and the MW probe pulse. Finally, the atoms fly towards the field-ionization detector $D$ outside the electrode structure. It resolves states in adjacent Rydberg manifolds, but does not resolve the relevant spin states.

All parameters of the experiment are independently measured or extracted from fits between the data and a numerical simulation of the experiment taking into account the complete level structure.

The Wigner function measurement is based on a complete reconstruction of the atomic state $\rho_t$ in the $n_g$ and $n_e$ manifolds. Measurements of the populations of the $|n_g,k\rangle$ and $|n_e,k\rangle$ ($k<6$) levels after a rotation of the spin states are used to fit $\rho_t$ using a Maximum Likelihood procedure. The final result is then projected on the spin state ladder. This approach leads to a direct calibration of the experimental imperfections: about 9.3\% of the population is spuriously transferred in levels outside the union of ${\cal H}_N$ and ${\cal H}_S$.

{\bf Acknowledgements: } We thank S. Pascazio and P. Facchi for many discussions and fruitful exchanges. We acknowledge funding by ANR under the project `QUSCO-INCA' and by the EU under the ERC project `DECLIC'.

{\bf Author Contributions: } A.S. and A.F. contributed equally. A.S., A.F., D.G., I.D., J.M.R., S.H., M.B. and S.G. contributed to the experimental set-up. A.S., A.F. and D.G. collected the data and analysed the results.  J.M.R., S.H., and M.B. supervised the research. S.G. led the experiment and performed the numerical simulations. All authors discussed the results and the manuscript. 

{\bf Author Information: } The authors declare no competing financial interests.

\clearpage


\section{Supplementary information}

\paragraph{Circular state preparation}

The experiment takes place in the electrode structure (vertical symmetry axis $Oz$) sketched in figure Extended Material 1. The experiment is cooled down to 1 K. A voltage $V_z$ applied across the $A$ and $B$ electrodes produces in the vicinity of the center $O$ of the structure a nearly homogeneous static electric field aligned with $Oz$. Four electrodes (1,2,3,4) surrounding the gap between $A$ and $B$ are used to generate RF or dc transverse fields.

The Rydberg atoms are excited in a thermal Rubidium atomic beam propagating along axis $Ox$. They are promoted from the $5s$ ground state to the $49f$ level using a resonant three-photon excitation ($5s\rightarrow 5p$ transition at 780 nm, $5p\rightarrow 5d$ at 776 nm, and $5d\rightarrow 49f$ at 1258 nm). The 780 nm and 776 nm laser beams are collinear and propagate in the horizontal plane at a 45$^\circ$ angle with respect to $Ox$. Due to Doppler effect, laser excitation addresses atoms with a velocity $v=254\pm4$~m/s along $Ox$. A  static voltage applied across electrodes 2 and 4 creates a 0.236 V/cm field along the propagation direction of the 780 and 776 nm lasers and defines the quantization axis during laser excitation. Both lasers are  $\sigma_+$-polarized with respect to this axis.

The $\pi$-polarized 1258 nm laser  beam is also horizontal and perpendicular to the other beams. The static field lifts the degeneracy between the $|m|$ sub-levels of $49f$. With the chosen laser polarizations, we prepare selectively the  $49f, m=2$ state. The 1258 nm laser is switched on for 1~\us s. This results in the preparation of a sample with 1.5 atoms on the average and a spatial extension of the order of 1 mm, much smaller than the size of the electrode structure (60 mm diameter).

After the laser excitation pulse, the transverse static field is adiabatically switched off in 1 $\mu$s, while the field along the quantization axis $Oz$ raises up to $2.55$ V/cm (linear Stark frequency 240 MHz for the $n=49$ manifold). We then apply a $\sigma_+$ RF field at 230 MHz with the electrode pair (1,2). The static field is then decreased to 2.31 V/cm (Stark frequency 218 MHz) in 1 $\mu$s. This procedure realizes an adiabatic passage on resonance with the transition between the lower levels of the 49 Stark manifold and transfers the atoms in the $49$ circular state with a high efficiency. The RF is then switched off and the electric field $F$ is set to the constant 2.35 V/cm value used for the main experimental sequence.

We finally apply a 0.9 \us s MW `purification' pulse, tuned to the two-photon transition from the circular 49 level towards the circular 51 state, which does not address the spurious population left in `elliptical' states of the 49 manifold with $m<48$. This ensures that only the state $\ket{n_e,0}$ is populated in the 51 manifold. 

The preparation sequence is completed 4.6 \us s after the end of the 1258 nm laser pulse.

\setcounter{figure}{0}
\renewcommand{\figurename}{EXTENDED FIGURE}

\begin{figure}
  \includegraphics[width=.8\linewidth]{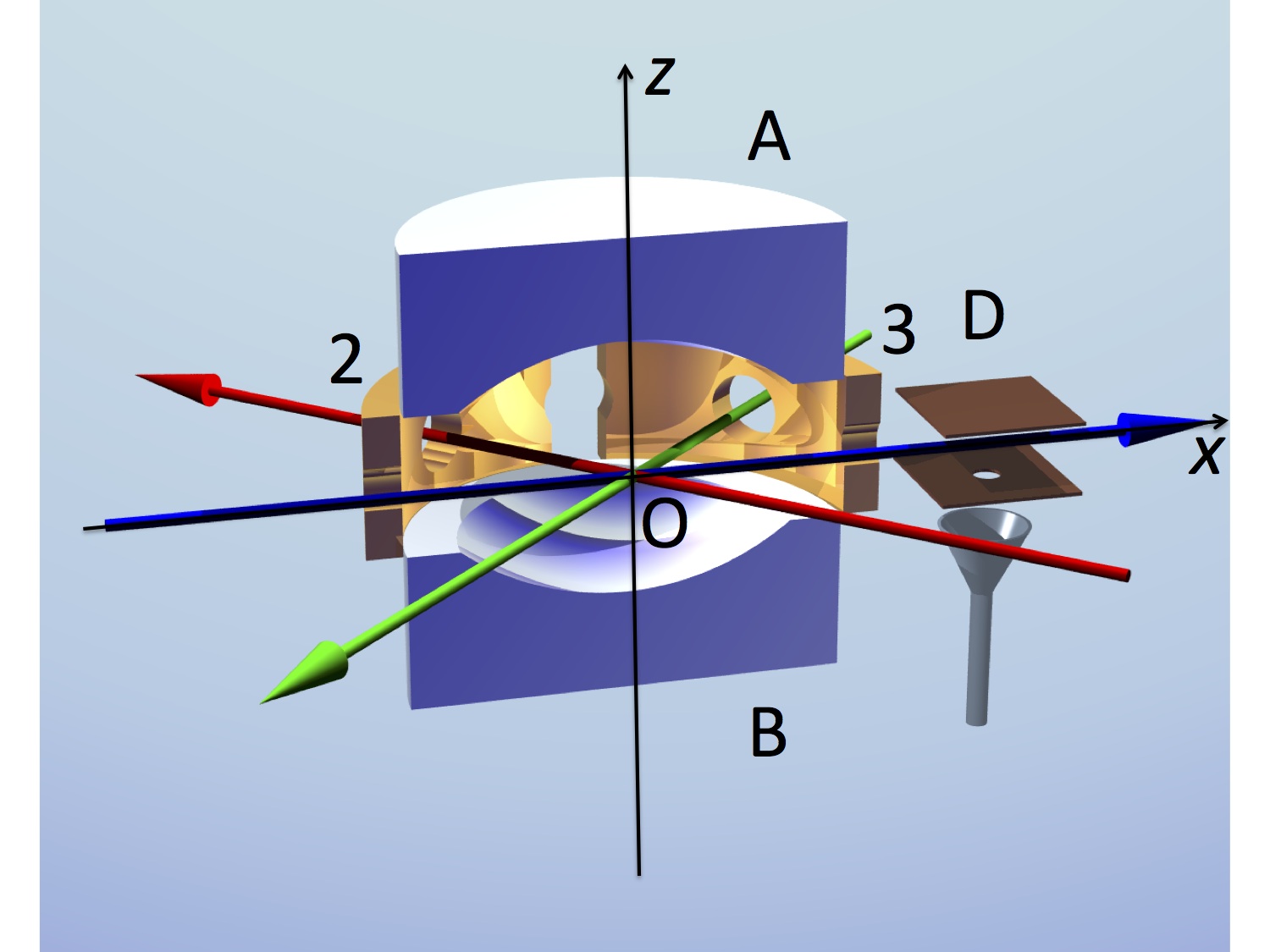}
  \caption{{\bf Schematic of the experiment} 
  The atoms are produced by excitation of a thermal Rubidium beam (blue arrow) propagating along axis $Ox$. Two mirror-shaped electrodes $A$ and $B$ (represented here cut by a vertical plane) produce the directing electric field $\mathbf{F}$ along $Oz$. The gap between $A$ and $B$ is surrounded by four independent electrodes (1, 2, 3 and 4), on which we apply RF signals to produce $\sigma_+$ fields with tunable phase and amplitude. Electrodes 1 and 4, not represented, are the mirror images of electrode 2 and 3 (in yellow) with respect to the $xOz$ plane.
 The laser excitation to the Rydberg states is performed using three laser beams that intersect in the center $O$ of the cavity. The 780 nm and 776 nm laser beams are collinear (red), the 1258 nm laser is sent perpendicular to the other beams (green). 
Once the atoms have left the electrode structure, they enter the field-ionization detector $D$.}
\end{figure}

\paragraph{Atomic state detection}

When they exit the electrode structure, the atoms enter the field-ionization detector $D$ \cite{ETQ} (Extended figure 1). We count the electrons produced by Rydberg levels ionization as a function of the field applied in $D$. This procedure resolves the $\ket{n_f,k}$, $\ket{n_e,k}$ and $\ket{n_g,k}$ states for a given $k$. Nevertheless, it does not resolve all relevant levels. In particular, $\ket{n_e,k}$ states with similar $k$ values ionize in nearly the same field.

In order to assess precisely the population $P(k_p,t_1)$ of $|n_e, k_p\rangle$, after the RF-induced QZD, we thus apply a 0.9 $\mu$s MW `probe' $\pi$-pulse, calibrated to transfer selectively and efficiently the population of $|n_e, k_p\rangle$ into $|n_f,k_p\rangle$. Due to the difference in the linear Stark frequencies between adjacent manifolds, transitions originating from other $|n_e, k\not =k_p\rangle$ states are detuned by $(k-k_p)\Delta$ with respect to the probe pulse, with $\Delta/2\pi=4.6$ MHz. This ensures the probe pulse state selectivity.

We then measure the population, $P_{n_f}(k_p)$ of $|n_f,k_p\rangle$. To that end, we count the average number of atoms detected at the ionization peak of $|n_f,k_p\rangle$ after the probe pulse. We subtract the number of atoms detected at the same field without the probe pulse, corresponding to the background produced by the ionization of $|n_e,k>k_p\rangle$ states. $P_{n_f}(k_p)$ is finally obtained by normalization to the average number of atoms initially prepared in the $51$ circular state.

The probe pulse has a finite transfer efficiency,  $\eta_{k_p} = P_{n_f}(k_p)/P(k_p,t_1)$. It is determined through the fit of the data presented in Fig.2~(a) with a numerical simulation of the experiment. This provides $\eta_{k_p}=\{0.90\pm0.02,0.90\pm0.01,0.90\pm0.01,0.90\pm0.01,0.87\pm0.02,0.84\pm0.02\}$ for $k_p=0\ldots 5$.

\paragraph{Dressed states and Zeno MW adiabatic switching-off}

In a proper interaction representation with respect to the Zeno MW drive, the dressed states are $|\pm\rangle=(|n_e,k_z\rangle\pm|n_g,k_z\rangle)/\sqrt 2$. To measure their population, we implement an adiabatic transfer of $\ket +$ and $\ket -$ onto $|n_e,k_z\rangle$ and $|n_g,k_z\rangle$ respectively at the end of the interaction with the RF field. We first decrease in 3~\us s the value of the electric field to detune the Zeno MW from the $\ket{n_g, k_z}\rightarrow\ket{n_e, k_z}$ transition. It sets the MW frequency exactly halfway between the $\ket{n_g, k_z}\rightarrow\ket{n_e, k_z}$ and $\ket{n_g, k_z+1}\rightarrow\ket{n_e, k_z+1}$ transitions. This correspond to a reduction of the amplitude of $F$ by a factor 0.90 (for $k_z=4$) or 0.92 (for $k_z=5$).  We then ramp down the amplitude of the microwave field and switch it off after 4.5 $\mu$s. This procedure maps with a high efficiency $\ket +$ onto  $|n_e,k_z\rangle$ and $\ket -$ onto  $|n_g,k_z\rangle$. Other levels $|n_e,k\not =k_z\rangle$ are only slightly mixed with the corresponding $n_g$ manifold levels and map with unit efficiency to the unperturbed states at the end of the adiabatic transfer procedure.

\paragraph{Experimental sequence}

The time origin of the sequence is chosen to be the start of the 1258 nm laser pulse. The preparation of the circular $51$ state ends at $t= 5.6$~$\mu$s. The Zeno MW is switched on at $t=6$~\us s. The first RF pulse of variable duration $t_1$, applied with the (1,2) electrode pair, always ends at $t=8$~\us s. For state reconstruction ($Q$ or $W$ function measurements), the QZD is immediately followed by a rotation of the spin produced by an intense RF pulse, applied on electrodes 3 and 4 during a time interval $t_2$ of 0.3~\us s at most. The Zeno MW adiabatic switching-off lasts from $t=9$~\us s to $t=16.5$~\us s: the electric field is reduced between $t=9$~\us s and $t=12$~\us s, before slowly decreasing the Zeno MW amplitude in 4.5 \us s. The electric field is then switched to $F=2.35$ V/cm at $t=17$~\us s, and the MW probe pulse is applied. Due to the stationary wave structure of the MW, the exact start time of the probe pulse is chosen to optimize the efficiency of the transfer rate $\eta_{k}$. Its duration is $0.9$ \us s. The atoms finally reach the detector $D$ around $t=232\ \mu$s.

\begin{figure}
  \includegraphics[width=.8\linewidth]{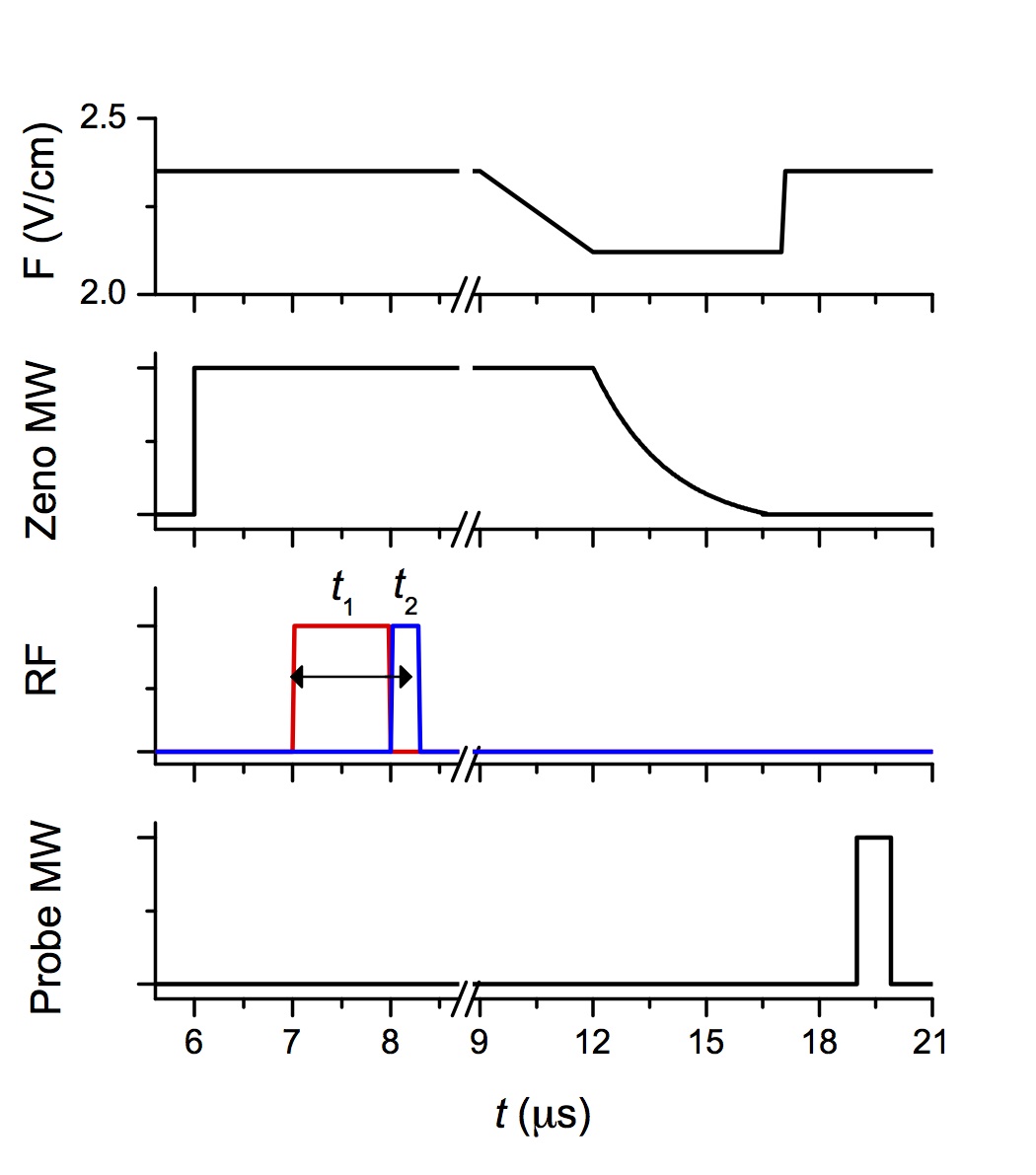}
  \caption{{\bf Timing of the experimental sequence} The Zeno MW is switched on at $t=6$ \us s with respect to the excitation laser pulse. The RF pulse of duration $t_1$ starts at an adjustable time between  $t=6.14$ \us s and $t=7.94$ \us s and ends at $t=8$~\us s (red). For the state reconstruction experiments, an intense RF pulse immediately follows for a duration $t_2$ (blue). At $t=9$~\us s, the electric field is reduced and the amplitude of the Zeno MW is then slowly decreased. The electric field is finally switched to $F=2.35$ V/cm at $t=17$~\us s, and the MW probe pulse is applied.}
\end{figure}

\paragraph{Calibration methods} 

The quality of the RF circular $\sigma_+$ polarization is of paramount importance in this experiment. To generate the $\sigma_+$ RF field, we optimize the phase and amplitude of the RF drives applied either on (1,2) or (3,4) electrodes through a procedure that will be described elsewhere. We measure the residual $\sigma_-$ polarization amplitude to be on the order 5 \% of the $\sigma_+$ field.

The linear Stark frequency is determined by the static field $F$ produced by electrodes $A$ and $B$. They are indeed spherical mirrors, ordinarily used in microwave cavity QED experiments. For a constant $V_z$ applied across them, the field varies with the atomic position in the structure. The resulting spatial variation of the Stark frequency along the atomic beam has been measured by MW spectroscopy and is compensated in the actual experiments by applying a time-varying $V_z$. With this procedure, $\omega_a/2\pi$ varies by less than 30 kHz during the duration of the RF pulses.

A precise calibration of the linear Stark frequency $\omega_a(F)$ as a function of $F$ is performed, for a few $F$ values, by Ramsey spectroscopy. Two short RF pulses at angular frequency $\omega_{rf}$ are applied on atoms in the circular $51$ state, separated by a time $T$. We measure the probability $P(0,T)$ to get the atom in the initial state as a function of $T$. The resulting $P(0,T)$ oscillates at the angular frequency $|\omega_a(F)-\omega_{rf}|$. The finite set of calibrated $\omega_a(F)$ values is interpolated to set the atom frequency 150~kHz above the RF one. This small detuning nearly compensates on average the light shifts produced by the Zeno MW on the $|n_e,k\not =k_z\rangle$ states.

The Zeno MW Rabi frequency, $\Omega_{mw}$, is inferred from a measurement of the Rabi splitting between the dressed states $\ket +$ and $\ket -$. We prepare the $\ket{n=49,k_z}$ state and probe the transitions towards  $\ket \pm$ with an auxiliary microwave field. We then switch off abruptly the Zeno MW and measure the probability to find the atom in $\ket{n_g,k_z}$ as a function of the frequency of the auxiliary field.

The Rabi frequency of the RF field inducing the QZD, $\Omega_{rf}$, is calibrated by the fit of the observations in Fig. 2(a) with a numerical simulation. A similar experiment calibrates the Rabi frequency for the intense RF rotation used for state reconstruction, $\Omega'_{rf}$.

The static electric field inhomogeneity across the atomic sample size and trajectory is estimated by rotating by an angle $\theta_0$ the $|n_e,0\rangle$ state by a first RF pulse. This creates a coherent spin state. We let this state evolve in the inhomogeneous electric field for an adjustable time delay $T$. We then measure its $Q(\theta,\phi)$ function as a function of $\phi$ at the latitude $\theta=\theta_0$. The frequency dispersion $\Delta\omega_a$ induced by field inhomogeneities leads to a phase-broadening of the $Q$-function.  Its measured Gaussian width, $\Delta\phi$,  is $\Delta\phi^2=\Delta\phi_0^2+\Delta\omega_a^2T^2$, where  $\Delta\phi_0$ is  the intrinsic width of the spin coherent state. Repeating the experiment for a few $T$ values, we get $\Delta\omega_a/2\pi= 174$~kHz. This inhomogeneity is the main source of decoherence for these experiments. It could be considerably reduced in a optimized electrode structure.

\paragraph{Numerical simulations}

Numerical simulations are performed in the complete Hilbert space of the atom, generated by all $|n_g,k\rangle$ and $\ket{n_e,k}$ states whose energy are calculated including second order Stark effect. $\omega_a$ is the frequency of the first step of the spin ladder. We compute the total Hamiltonian, including the RF and MW Zeno drives in a proper interaction representation. The RF drive is represented by the interaction potential $V$ [Eq. (1)] with constant amplitude. The permanent MW drive is approximated by a constant Rabi frequency coupling between $|n_g,k\rangle$ and $\ket{n_e,k}$ for all $k$ values. To get the evolution operator $U_{QZD}$ representing the Zeno dynamics, we numerically exponentiate this Hamiltonian with MATLAB during the effective time interval $t_1-t_1^0$  where $t_1^0= 69\pm 1$ ns is a small offset representing the finite rise time of the actual RF field amplitude. This offset is extracted from a fit of the data presented in Fig. 2(a). 

For the quantum state reconstruction procedure, the preparation pulse of duration $t_1$ is nearly immediately followed by the rotation pulse of duration $t_2$. This is modeled in the simulation by two ideal square pulses of effective duration $t_1-t_1^0$ and $t_2-t_2^0$, separated by a time interval $t_i$ during which the RF field amplitude is set at zero. The time $t_i=30$~ns is an adjustable parameter that models the transient regime due to the actual shape of the pulses. The time $t_2^0=74\pm1$~ns is extracted from the fit of an auxiliary experiment.

Finally, we take into account the static electric field inhomogeneity by simulating the experiment for different value of $F^\prime$, and averaging the results with a Gaussian weight $p(F^\prime)$ centered around $F$.

\paragraph{MaxLike analysis}

We reconstruct the complete atomic density operator $\rho_t$ in the basis of the $|n_e,k\rangle$ and $|n_g,k\rangle$ states for all $k$ values using the MaxLike algorithm \cite{Lvovsky2004}. This provides full information on the spin state at the time of the cat creation (see Fig. 4). We apply an intense rotation RF pulse, with adjustable phase and rotation angle, and  adiabatically switch off the Zeno MW. This complex procedure is represented by a unitary operator $U$ acting on the full atomic state. The operator $U$ is numerically obtained through the exhaustive numerical simulation of the procedure. We then measure the populations in the  $|n_e,k\rangle$ and $|n_g,k\rangle$ states  for $k<6$ and fit by a Maximum likelihood procedure the state $\rho_t$ to these observations. We finally get $\rho$ by projecting $\rho_t$ on the spin state ladder. We normalize the result by a coefficient $1/0.91$ corresponding to the spurious leaks out of the spin Hilbert space. We check in the procedure that the spurious population in the dressed state $\ket -$ is lower than 1.5\% at the reconstruction time. The Wigner function is computed according to Ref \cite{Dowling1994}. Since the spin states with $k\ge16$ have a negligible population (less than 0.7\%), dominated by noise, we truncate $\rho$ to the first 16 spin states before computing $W$.

\end{document}